\documentclass[twocolumn,times]{aastex63}
\usepackage{subfigure}
\usepackage{gensymb}
\usepackage{tablefootnote}
\usepackage{enumitem}
\definecolor{malachite}{rgb}{0.04, 0.85, 0.32}

\usepackage{soul,xcolor}
\setstcolor{red}

\submitjournal{ApJ}
\begin{document}

\shorttitle{The Shape of AGN-Driven Winds in the Seyfert Galaxy NGC 3516}

%\watermark{text}
%\setwatermarkfontsize{dimension}
%\graphicspath{{./}{figures/}}

%\begin{document}

\title{The Shape of AGN-Driven Winds in the Seyfert Galaxy NGC 3516}

\author[0000-0002-2713-8857]{Jacob Tutterow}
\affiliation{Department of Physics and Astronomy, Georgia State University, 25 Park Place, Atlanta, GA 30303, USA}

\author[0000-0002-7130-7099]{Nicolas Ferree}
\affiliation{Department of Physics, Stanford University, 382 Via Pueblo Mall, Stanford, CA 94305, USA}

\author[0000-0002-6465-3639]{D. Michael Crenshaw}
\affiliation{Department of Physics and Astronomy, Georgia State University, 25 Park Place, Atlanta, GA 30303, USA}

\author[0000-0001-7238-7062]{Julia Falcone}
\affiliation{Department of Physics and Astronomy, Georgia State University, 25 Park Place, Atlanta, GA 30303, USA}

\author[0009-0005-3001-9989]{Maura Kathleen Shea}
\affiliation{Department of Physics and Astronomy, Georgia State University, 25 Park Place, Atlanta, GA 30303, USA}

\author[0000-0002-3365-8875]{Travis C. Fischer}
\affiliation{AURA for ESA, Space Telescope Science Institute, 3700 San Martin Drive, Baltimore, MD 21218, USA}

\author[0000-0001-8658-2723]{Beena Meena}
\affiliation{Space Telescope Science Institute, 3700 San Martin Drive, Baltimore, MD 21218, USA}

\author[0000-0002-4917-7873]{Mitchell Revalski}
\affiliation{Space Telescope Science Institute, 3700 San Martin Drive, Baltimore, MD 21218, USA}

\author[0009-0009-4283-3311]{Kesha Patel}
\affiliation{Department of Physics, Emory University, 400 Dowman Drive, Atlanta, GA 30322, USA}

\author[0009-0005-2145-4647]{Madeline Davis}
\affiliation{Department of Physics and Astronomy, Georgia State University, 25 Park Place, Atlanta, GA 30303, USA}

%\correspondingauthor{Jacob Tutterow}
%\email{jtutterow1@gsu.edu}

%\author[0000-0000-0000-0000]{...}
%\affiliation{Department of Physics and Astronomy, Georgia State University, Atlanta, GA 30302-4106, USA}

%\collaboration{1}{(...)}
    
\begin{abstract}

Active galactic nuclei (AGN) are known to drive ionized gas into their host galaxies, which may affect the evolution of both the central supermassive~black~holes (SMBHs) and their hosts. In the case of NGC~3516, a nearby Seyfert 1 galaxy, these AGN winds have historically proven difficult to disentangle from galactic rotation. Using long-slit spectroscopy at multiple position angles from Hubble Space Telescope’s Space Telescope Imaging Spectrograph (STIS) and Apache Point Observatory’s Kitt Peak Ohio State Multi-Object Spectrograph (KOSMOS), we separate these kinematic components by fitting multiple Gaussians to the H$\alpha$, [N II], H$\beta$, and [O III] emission lines along the slits. We present a biconical outflow model that agrees well with the observed kinematics of the outflowing gas in the narrow-line region (NLR). Our results indicate that the structure of the [O~III] emission is explained by dusty gas spirals in the galactic disk that are illuminated by the ionizing bicone, which is viewed along one edge, resulting in the complex nuclear kinematics. Our view into the bicone edge is consistent with the multiple, deep components of ionized absorption lines seen in UV and X-ray spectra of NGC~3516. The observed turnover in the velocity of the NLR clouds matches that from a simple dynamical model of radiative acceleration by the AGN and gravitational deceleration by the AGN and galaxy, indicating they are the principal forces at work on the gas clouds. Finally, the model launch radii indicate that the outflowing clouds originate primarily from the inner dusty spirals near the AGN.

\end{abstract}


%\keywords{galaxies: active — galaxies: individual (NGC 3227, NGC 3226) ― galaxies: Seyfert ― ISM: jets and outflows}

\section{Introduction}

Supermassive black holes (SMBHs), with masses ranging between $10^6 - 10^9$ M$_\odot$, lie at the centers of most galaxies. Those SMBHs that are actively accreting material are luminous enough to be classified as active galactic nuclei (AGN). Radiation from the accretion disk ionizes the surrounding broad-line region (BLR) and narrow-line region (NLR) clouds of gas, whose emission line kinematics provide clues to the dynamical forces at work near the AGN. An obscuring torus of dusty gas lies between these two regions, leading to significant spectral differences depending on viewing angle. AGN generally fall into two types, with type 1 AGN being unobscured in our line of sight (LOS), and type 2 having nuclei obscured by the torus according to the commonly accepted unified model \citep{antonucci85}. Along with orientation of the AGN, they are also classified based on their bolometric luminosity. This study focuses on Seyfert galaxies, which range in luminosity from L$_{bol} = 10^{43} - 10^{45}$ erg s${^{-1}}$ and contribute significantly to the observed emission.

As AGN emit powerful amounts of radiation, the intense radiation pressure can ionize and accelerate ambient gas (and dust) into the host galaxy to generate outflows or ``AGN winds'' \citep{fischer17, fischer18, revalski21}. The AGN winds can also further impact the surrounding gas and dust. The radiation driving and/or subsequent wind interaction can provide ``AGN feedback'' into the galaxy \citep{fabian12}. For example, AGN winds can potentially either kick-start \citep{silk13, shin19} or quench star formation \citep{springel05, bower06, canodiaz12}. The most common view is that AGN feedback is thought to evacuate the galaxy's nucleus of gas, starving the AGN and leaving the SMBH in a quiescent phase \citep{fabian12}. Eventually the gas may repopulate the nucleus (if it does not escape the galaxy), causing accretion onto the SMBH again, leading to a cycle that can dramatically affect the galaxy's evolution \citep{booth09, angles17}.

From the unified model, the ionized gas, and hence the ionized AGN winds are expected to follow a biconical geometry, following the opening angles of the obscuring torus \citep{antonucci85}. High velocity (FWHM = 800 $-$ 8000 km s$^{-1}$) clouds of dense gas (n$_{H} = 10^{8} - 10^{12}$ cm$^{-3}$) close to the accretion disk form the BLR, seen in spectra as broad components of emission lines. Farther out from the AGN, slower moving (FWHM $\leq$ 2000 km s$^{-1}$), more diffuse clouds of gas (n$_{H} = 10^{2} - 10^{6}$ cm$^{-3}$) form the NLR. These are seen in narrow components of emission lines in spectra of AGN. While the BLR is only a few light days in size around the SMBH, the NLR can extend for kiloparsecs throughout the galaxy. The NLR kinematics have been spatially resolved and accurately modeled in a number of nearby Seyfert galaxies \citep{crenshawkraemer00, fischer13, meena23, falcone24}.

In this work we study another nearby Seyfert galaxy, NGC~3516.
NGC 3516 is one of the original Seyfert galaxies discovered \citep{seyfert43}, and is also the first observed AGN to change spectral features, as \cite{Andrillat1968} compared 1967 measurements to Seyfert's original 1943 measurements and claim to have found Balmer line enhancement and a decrease in [O~III] intensity \citep{souffrin73}. It has since been classified as a changing look AGN \citep{shapovalova19}, although it is usually seen as a Seyfert 1 galaxy. It is a barred lenticular (R)SB0$^{0}$?(s) galaxy \citep{deVaucouleurs91} that is relatively nearby (z = 0.008836, D $\approx$ 37.1 Mpc, R $\approx$ 200 pc arcsec$^{-1}$; \citealp{goad&gallagher87, keel96}). It has a prominent Z-shaped NLR seen in [O~III] emission that is ionized by the AGN, that spans $\sim$13\farcs6 ($\sim$2330 pc; \citealp{schmitt03, ferruit98}), and its variability in continuum emission, broad emission lines, and UV/X-ray absorption lines has been extensively studied \citep{goad99, markowitz03, dunn18, Oknyansky21, ilic24}. This variability has been attributed to changes in the spectral energy distribution (SED) and/or flux, which has changed the ionization of the warm-absorber outflows \citep{mehdipour22}.
% Eddington luminosity = ~5.37490193691 * 10^45

% bolemtric lum = 9.8(L_5100) = 

Determining the extent and kinematics of the NLR outflows in NGC 3516 has produced ambiguous results \citep{fischer13}. The main reasons for this difficulty were the non-unique solutions for the parameters of the bicone, and inconclusive evidence for symmetrical NLR outflows. \cite{fischer13} used two Hubble Space Telescope (HST) Space Telescope Imaging Spectrograph (STIS) observations for their study, and found it difficult to separate the rotation of the galaxy from outflows in the nucleus. In this paper, we use more long-slit spectra with varying position angles and slit widths to gain a deeper understanding of the AGN wind kinematics through more accurate modeling of the outflowing bicone. %Awesome!!

 \label{sec:intro}
\section{Observations}

\begin{figure*}

\centering  
{\includegraphics[width=0.65\linewidth]{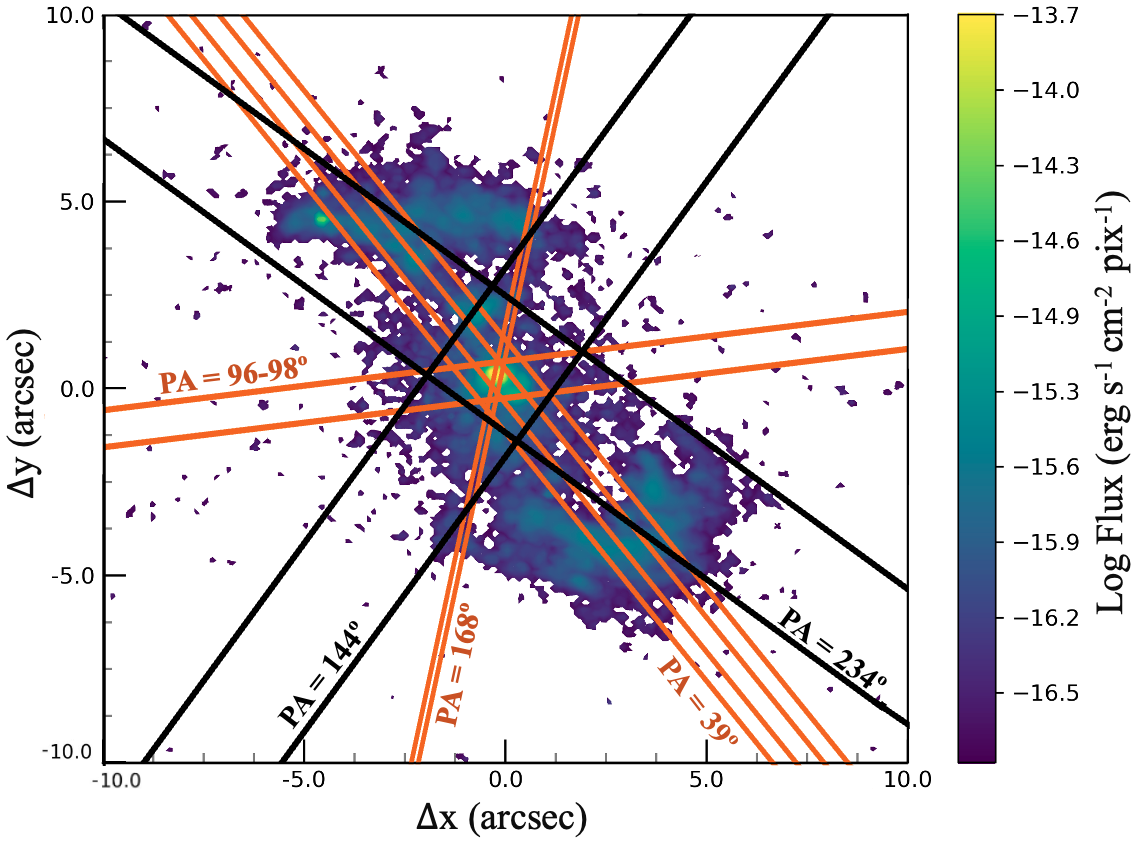}}

 \caption{%Position angles and slit sizes of
The long-slit positions used in this study, overlayed on top of a continuum subtracted [O~III] image (adapted from \citealp{schmitt03}) of NGC 3516. The black slits are from KOSMOS, and the orange slits are from STIS.\label{fig: positionangles}}

\end{figure*}

We use archival data from the Hubble Space Telescope's Space Telescope Imaging Spectrograph (STIS), along with new observations made with the Astrophysical Research Consortium (ARC) 3.5-meter telescope at Apache Point Observatory (APO), using the Kitt Peak Ohio State Multi-object Spectrograph (KOSMOS). The specifications of the observations used are given in Table~\ref{table:obs}, and a visual representation of the position angles (PAs) and slit sizes are given in Figure~\ref{fig: positionangles}. Different PAs are used to capture different parts of the NLR and galaxy. These observations were primarily used to measure two sets of emission lines. The first is H$\beta$ $\lambda$4861~\r{A} and [O III] $\lambda \lambda$4959, 5007~\r{A}, and the second is H$\alpha$ $\lambda$6563~\r{A} and [N II] $\lambda \lambda$ 6548, 6583~\r{A}.

% \begin{figure}
% \centering
% \includegraphics[width=1\linewidth]{fig/ngc3516_slitpos.png}%
% {%
%   \caption{Position angles and slit sizes of observations used in this study, overlayed on top of an image of a continuum subtracted [OIII] image of NGC 3516, taken from \cite{schmitt03}. The black slits are from KOSMOS, and the orange slits are from STIS.}
%   \label{fig: positionangles}
% }

% \end{figure}

% \begin{figure*}
% \centering  
% {\includegraphics[width=0.4\linewidth]{fig/positionangles.png}}
%  \caption{Position angles and slit sizes of observations used in this study, overlayed on top of an [OIII] image of the nucleus of NGC 3516.}
% \label{positionangles}
% \end{figure*}

\begin{table*}
%\centering
%\footnotesize
\begin{center}
%\begin{tabular} 
%{|c c c c c|} 
  \begin{tabular}{|l c c c c c c c c c c c|} \hline

% \hline

% Instru-	& Position & Offset & Grating & Slit Width \\
%	ment& Angle & (arcseconds) & & (arcseconds) \\
 Instru-	& Program &Date 	& Grating/	& \# of &Total & Wavelength & Spectral	& Slit & Spatial  &	Position  & Spatial \\
 
 ment & ID & (UT)  & Grism	    & Obs.  &Exposure &  Range 	& Dispersion& Width & Scale  &Angle & Offset\\

& & & & 	&  Time (s)& (\AA) &  (\AA/pix) & ($''$) &($''$/pix)& (deg) & ($''$ NW)\\

(1) &(2) &(3) &(4) 	& (5)& (6) &  (7) & (8) &(9)& (10) & (11) &(12)\\

 \hline
STIS & 8340 &2011 Jan 18 & G430M &3& 2154 & $4950-5236$ & 0.28 & 0.1& 0.051 & 168\degree & 0\\
STIS& 8055 &2000 Jun 18 & G750M &1& 1332 & $6248-6912$ & 0.56 &0.2& 0.051 & 39\degree & 0 \\
STIS & 8055 &2000 Jun 18 & G750M &1& 1200 & $6248-6912$ & 0.56 &0.2& 0.051 & 39\degree & 0.3\\
STIS & 8055 &2000 Jun 18 & G750M &1& 1200 & $6248-6912$ & 0.56 &0.2& 0.051 & 39\degree & -0.3\\
STIS & 8055 &2000 Jun 18 & G430L &1& 600 & $2900-5700$ & 2.73 &0.2& 0.051 & 39\degree & 0\\
STIS & 7355 &1998 Apr 14-16 & G430L &38& 128200 & $2900-5700$ & 2.73 &0.5& 0.051 & 96\degree -- 98\degree & 0\\
KOSMOS & GS01 &2022 Feb 24 & Blue-High &3& 2700 & $4150-7050$ & 0.71 &2& 0.257 & 144\degree & 0\\
KOSMOS & GS01 &2024 May 9 & Blue-High &3& 2700 & $4150-7050$ & 0.71 &2& 0.257 & 234\degree & 0\\

%  Instru-	& Program	& Date 	& Grating/	& \# of &Total & Wavelength & Spectral	& Slit & Spatial  &	Position  & Spatial \\
 
%  ment.& ID & (UT)  & Grism	    & Obs.  &Exposure &  Range 	& Dispersion& Size & Scale  &Angle & Offset\\

% & & & & Time (s)	&  (\AA)&   (\AA/pix)	&  ($''$/pix) & (deg) &($''$ NW)& & &\\

%  \hline
% STIS & 8340 & 2011 Jan 18 & G430M &3& 2154 & $4950-5236$ & 0.28 & 0.1& 0.051 & 168\degree & 0\\
% STIS & 	8055 & 2000 Jun 18 & G750M &1& 1332 & $6248-6912$ & 0.56 &0.2& 0.051 & 39\degree & 0 \\
% STIS & 8055 & 2000 Jun 18 & G750M &1& 1200 & $6248-6912$ & 0.56 &0.2& 0.051 & 39\degree & 0.3\\
% STIS & 	8055 & 2000 Jun 18 & G750M &1& 1200 & $6248-6912$ & 0.56 &0.2& 0.051 & 39\degree & -0.3\\
% STIS & 8055 & 2000 Jun 18 & G430L &1& 600 & $2900-5700$ & 2.73 &0.2& 0.051 & 39\degree & 0\\
% STIS & 7355 & 1998 Apr 14-16 & G430L &38& 128200 & $2900-5700$ & 2.73 &0.5& 0.051 & 96\degree -- 98\degree & 0\\
% KOSMOS & GS01 & 2022 Feb 24 & Blue-High &3& 2700 & $4150-7050$ & 0.71 &2& 0.257 & 144\degree & 0\\
% KOSMOS & GS01 & 2024 May 9 & Blue-High &3& 2700 & $4150-7050$ & 0.71 &2& 0.257 & 234\degree & 0\\

 \hline

\end{tabular}
\end{center}
\caption{HST and APO observations of NGC 3516. The columns list (1) instrument used, (2) program ID from HST/APO, (3) date of observation, (4) grating or grism used, (5) number of observations, and (6) total exposure time of all observations, (7) wavelength range of grating or grism, (8) spectral dispersion of the spectrograph, (9) width of the slit of the spectrograph, (10) spatial scale of spectrograph, (11) position angle of spectrograph, and (12) offset of the slit position relative to the nucleus of NGC 3516.} 
\label{table:obs}
\end{table*}

\subsection{STIS}

We used archival STIS spectra to understand the small-scale (circum-nuclear) kinematics of NGC 3516, as they have high angular resolution along the slit (0\farcs1) and a small slit width (0\farcs1 -- 0\farcs5 arcseconds, 20 -- 100 pc). We obtained the reduced spectroscopic data from the Mikulski Archive at the Space Telescope Science Institute. We performed additional calibrations to remove cosmic ray hits and hot pixels, and combined spectra at the same approximate positions. In particular, the multiple spectra obtained at PA $=$ 96$\arcdeg$ -- 98$\arcdeg$ were averaged together. These exposures were from an intensive variability monitoring program, described in detail in \cite{edelson00}.

Four STIS observations have a PA of 39\degree. One of these is centered on the nucleus with the G430L grating, and three are taken with the G750M grating, with one centered on the nucleus and two offset from the nucleus at $\pm$0\farcs3. These spectra lie close to the major axis of the bicone, and intersect nearly the full extent of the bright [O~III] emission. STIS spectra at the other PAs cover primarily the nuclear region. The G430L and G430M grating spectra were used to measure the H$\beta$ and [O~III] emission lines, and the G750M grating spectra were used to measure the H$\alpha$ and [N~II] lines.
The spectral resolutions are approximately 9.0 \AA, 0.56 \AA, and 1.8 \AA\ (FWHM) for the G430L, G430M, and G750M gratings, respectively.

\subsection{KOSMOS}
To probe the kinematics of NGC 3516 on a wider scale, we obtained KOSMOS spectra to supplement the STIS data. All KOSMOS observations were taken with the blue grism setting, providing a wavelength range of 4150$-$7050 \r{A}. This range is enough to cover both the H$\beta$ and H$\alpha$ regions. Three 15 minute exposures were taken per PA, and stacked to give a total exposure time of 45 minutes. We reduced these observations using standard IRAF routines, including flat-fielding, bias subtraction and flux calibrations using standard stars from the \cite{oke90} catalog. We used arc lamps on the telescope for wavelength calibrations, which were taken at each telescope position. Additional calibrations were performed in the Interactive Data Language (IDL) to correct for the tilts of the slits with respect to the cross-dispersion directions, and to subtract night sky lines.
The spectral resolution of the KOSMOS spectra is approximately 5.1~\AA. The seeing during the KOSMOS observations was $\sim$1\farcs5 on 2022 February 24 and $\sim$0\farcs9 on 2024 May 9.

We observed two position angles, covering the major (PA = 234$\degree$) and minor (PA = 144$\degree$) photometric axes of the host galaxy \citep{schmitt03}. These were chosen to measure rotational motions of the galaxy, along with the extent of the outflows near the nucleus. \label{sec:methods}
\section{Spectral Analysis \& Kinematics}

\begin{figure}[t]
%\centering  
{\includegraphics[width=1.00\linewidth]{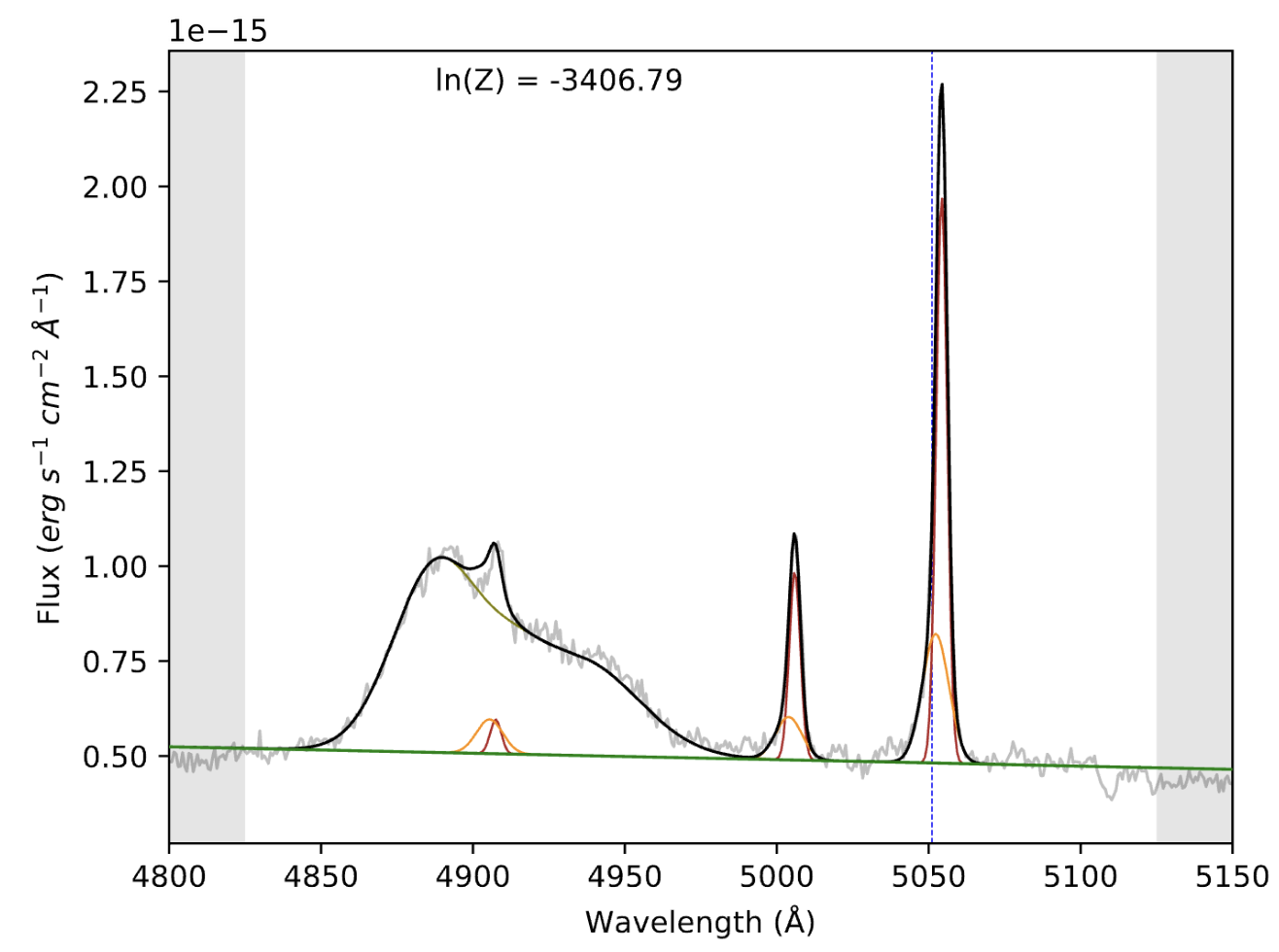}}
 \caption{An example of a multi-component Gaussian fit using BEAT, fitting the H$\beta$ and [O~III] emission lines in our KOSMOS observation of PA = 234$\degree$, at the nucleus of NGC~3516. The black curve is the overall fit, the brown-colored curve is the broad component fit, and the orange and red curves are narrow component fits. The gray shaded regions are regions BEAT uses to fit the continuum flux, with the continuum fit shown in green. The blue dotted line shows the position of the [O~III] $\lambda$ 5007 \AA\ emission line at the systemic redshift of NGC~3516. Z is the evidence value of the fit, as described by \cite{fischer17}.}
\label{fig:beatexample}
\end{figure}

%\subsection{Gaussian Line Fitting}
The long-slit spectra data allows us to characterize the kinematics of ionized gas as a function of distance from the nucleus. Using the Bayesian Evidence Analysis Tool (BEAT; \citealp{fischer17, falcone24}), we fit multi-component Gaussian profiles to the emission lines to separate the kinematic components at each position along the slit. The BEAT algorithm determines the number of significant Gaussian components needed to fit the emission lines in each wavelength region and returns the line centroid, full width at half maximum (FWHM), and emission-line flux for each component.
At each position, typically near the nucleus, we allow BEAT to fit up to 3 narrow components to the spectral line.
Tests were done with BEAT that allowed up to 5 Gaussian components;
however, there were never any fits that required more than 3 components.
In this study we are primarily concerned with the line centroid, from which we calculate the velocities of the gas components in the rest frame of the galaxy. An example fit from BEAT is shown in Figure~\ref{fig:beatexample}. 

Seyfert 1 galaxies, like NGC 3516, have both broad and narrow components for the permitted emission lines. As the broad component is unresolved in our observations, it is considered to be a point source and therefore only varies in flux along the slit according the point-spread function (PSF). To model the broad component of emission in BEAT, we extract the 1D spectrum from the nucleus and fit the broad-line profile with multiple Gaussians (attaching no significance to individual components). This shape is kept constant, and the profile only varies in flux along the 2D spectra to match the PSF. At each position along the silt, we fit the scaled broad-line profile, continuum on either side, and narrow-line emission using BEAT to determine the number of statistically significant narrow components.
The NLR kinematics from our STIS observations are given in Figure~\ref{fig: STIS}, and the NLR kinematics from our KOSMOS observations are given in Figure~\ref{fig: KOSMOSobs}.

\subsection{HST STIS Kinematics}
The STIS spectra allow us to probe small scales near the SMBH, as shown in Figure \ref{fig: STIS}. The available STIS position angles were chosen by the original observers and do not exactly match the major and minor axes of the host galaxy or the bicone axis of NGC 3516. Nevertheless, the slit positioned over the nucleus at a PA of 39$\degree$ shows clear evidence for components with both blueshifted and redshifted velocities up to 700 km s$^{-1}$ within $\pm$0\farcs1~($\sim$20 pc) of the nucleus, in both the [O~III] and H$\alpha$ lines, indicating outflow.
The offset slit positions containing H$\alpha$ at this PA do not show these high velocities, further indicating they are confined to the nucleus. Most of the velocities in all three slit positions at PA $=$ 39$\arcdeg$ are largely consistent with a rotation pattern spanning
$\pm$6$\arcsec$ from the central SMBH,
with the exception of a few points that will be addressed later.
The high velocity points, and numerous others within $\pm$0\farcs5 of the nucleus, show large FWHM, which are further possible indicators of outflow or at least kinematically disturbed gas \citep{fischer18}.

The [O~III] kinematics within $\pm$0\farcs1 of the nucleus at PA $=$ 168\arcdeg\ also show evidence of outflows at blueshifted radial velocities of $-$450 km s$^{-1}$. The [O~III] kinematics at PA $=$ 96\arcdeg\ -- 98\arcdeg\ show non-rotational motion, with mostly blueshifted velocities on either side of the nucleus, suggesting contributions from outflows.

\begin{figure*}
\centering  
{\includegraphics[width=0.42\linewidth]{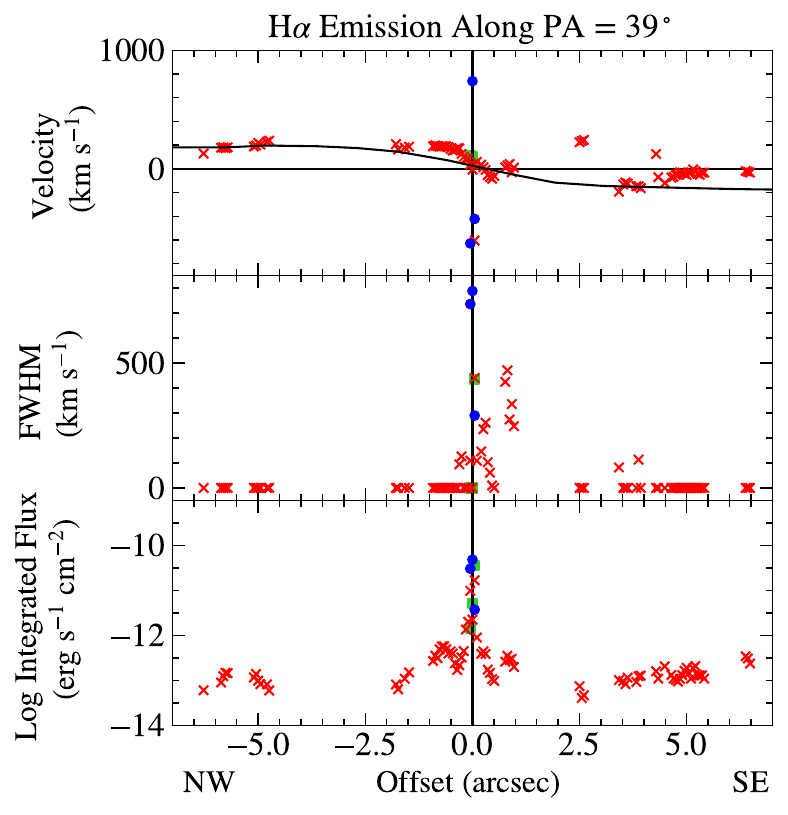}}
{\includegraphics[width=0.42\linewidth]{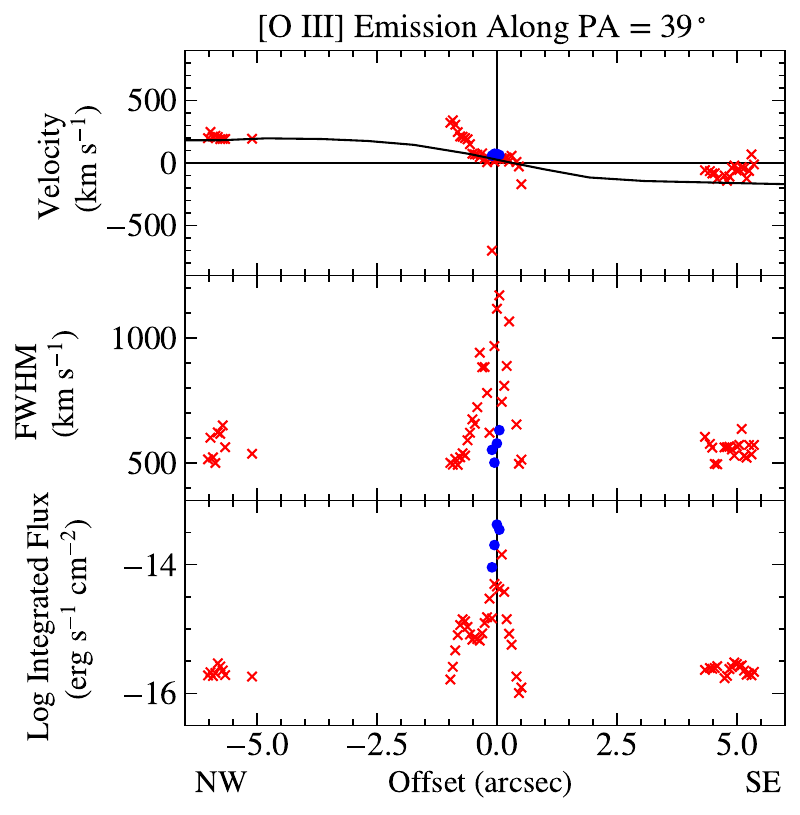}}
{\includegraphics[width=0.42\linewidth]{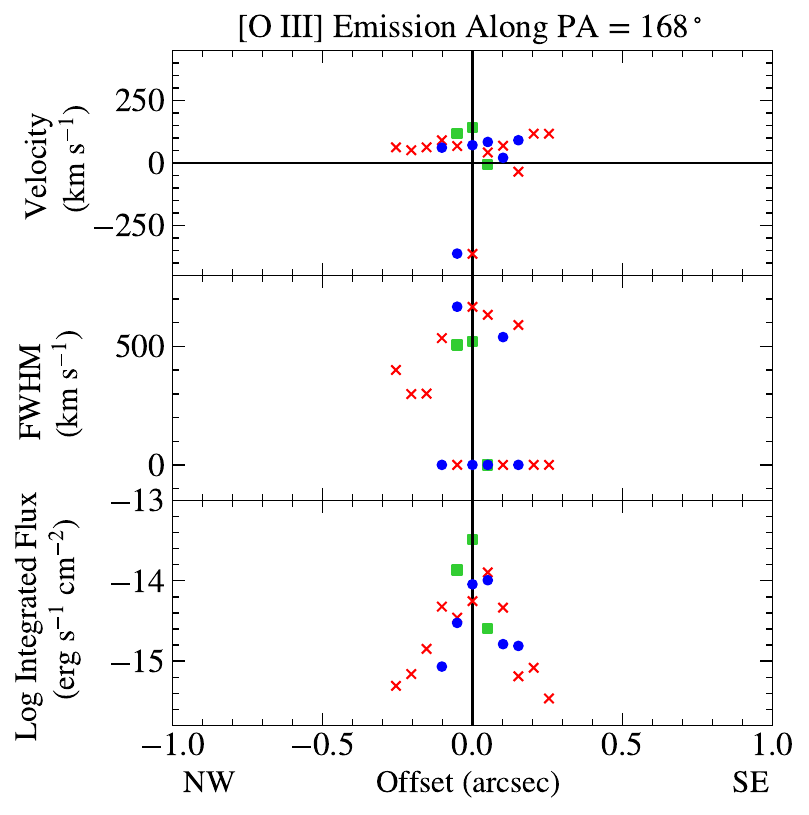}}
{\includegraphics[width=0.42\linewidth]{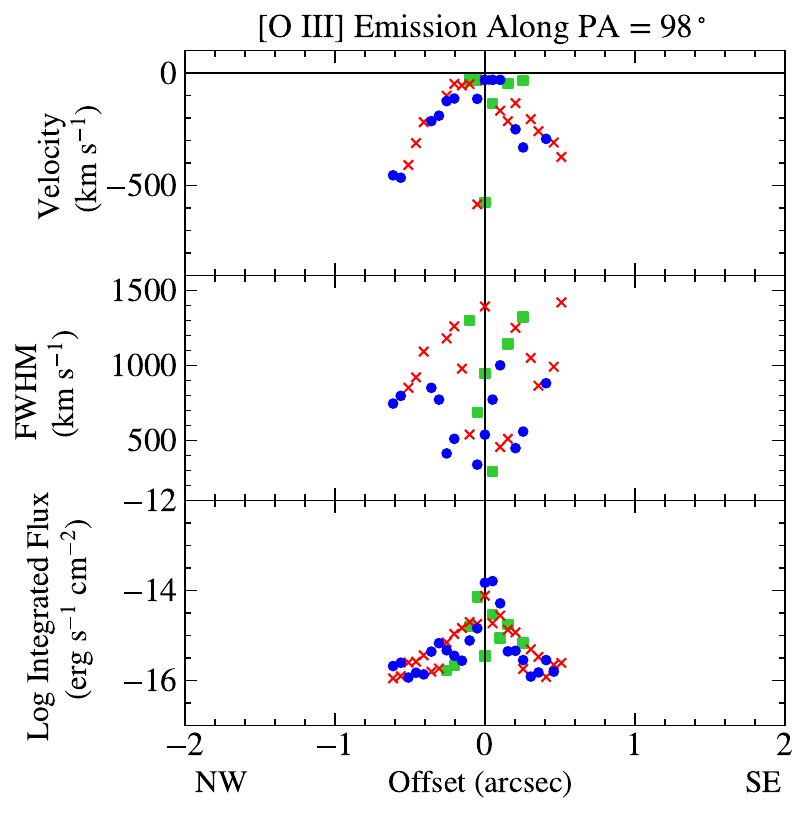}}
{\includegraphics[width=0.435\linewidth]{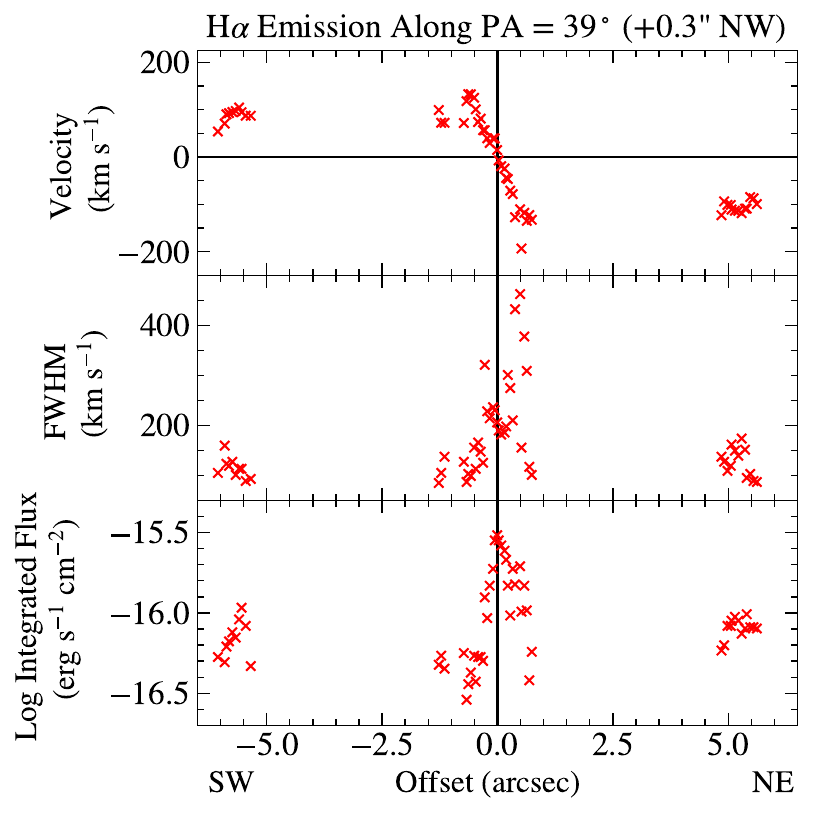}}
{\includegraphics[width=0.435\linewidth]{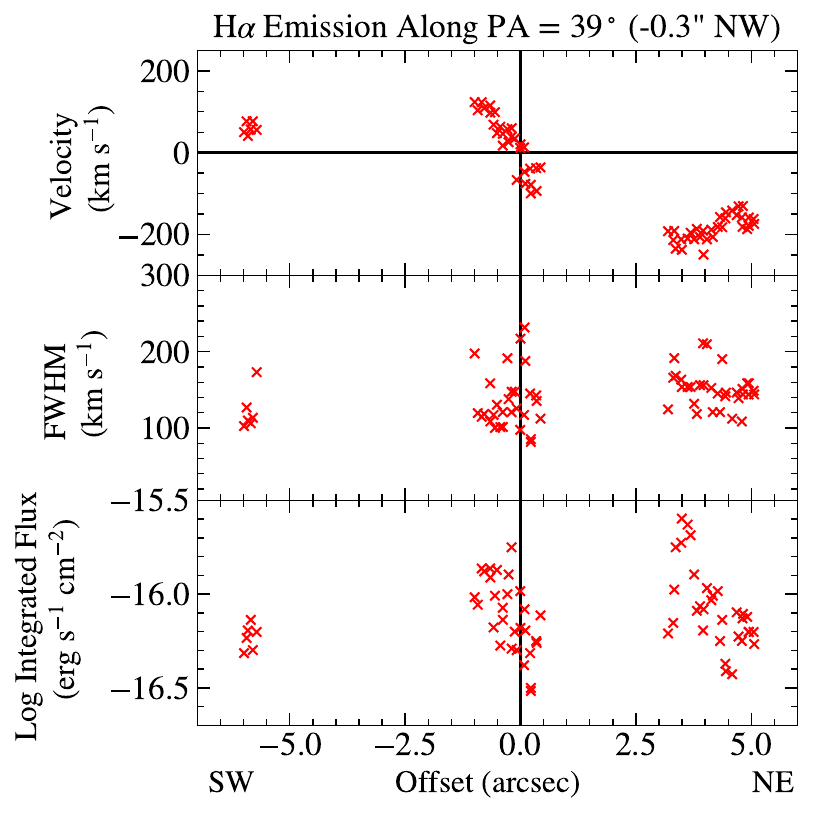}}
\caption{STIS observations of [O~III] and H$\alpha$ kinematics for various position angles. The top panels show the velocity, middle panels show the full-width at half maximum, and the bottom panels show the integrated flux. The different colors and shapes denote different components of the multi-component Gaussian fits. The components are sorted at each position according to increasing velocity. The black line in the H$\alpha$ kinematics close to the major axis is the rotation curve of NGC 3516, from \cite{Cherepashchuk2010}.}
\label{fig: STIS}
\end{figure*}

\begin{figure*}

\centering  
{\includegraphics[width=0.45\linewidth]{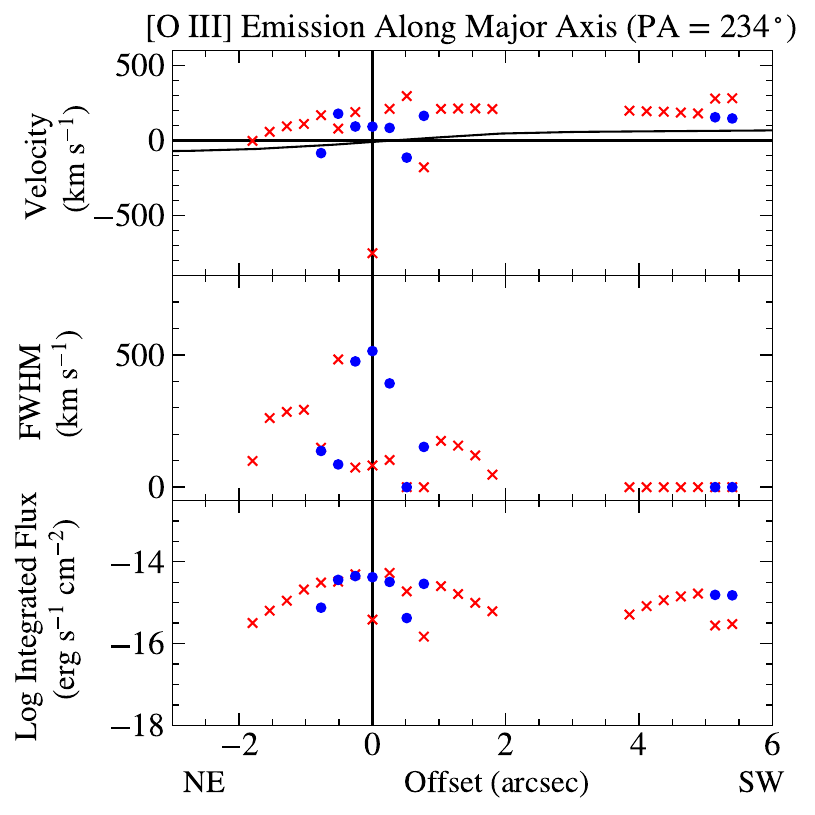}}
{\includegraphics[width=0.425\linewidth]{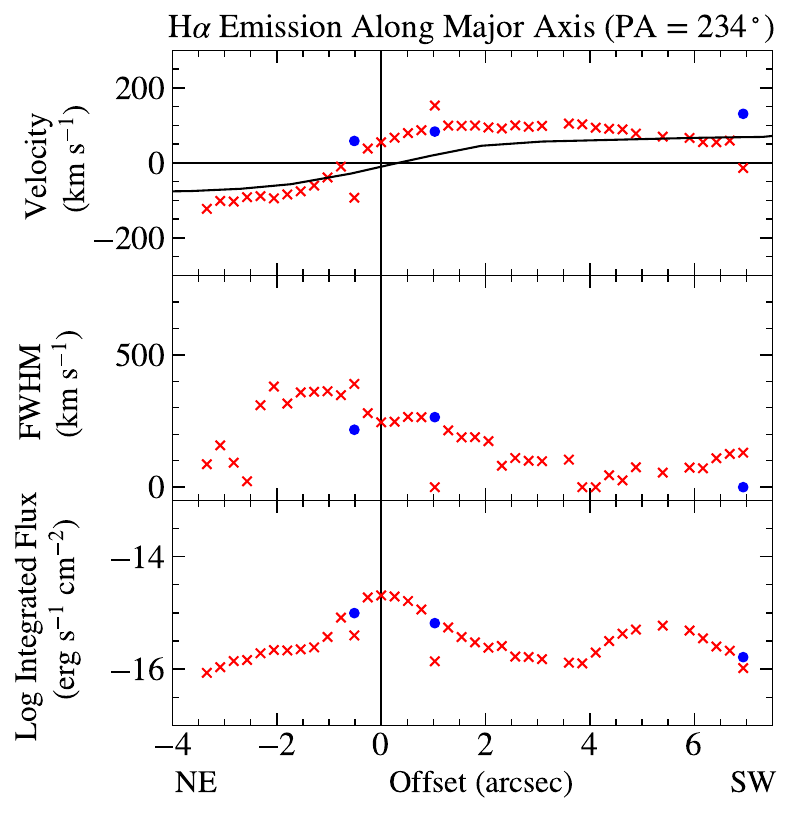}}
{\includegraphics[width=0.45\linewidth]{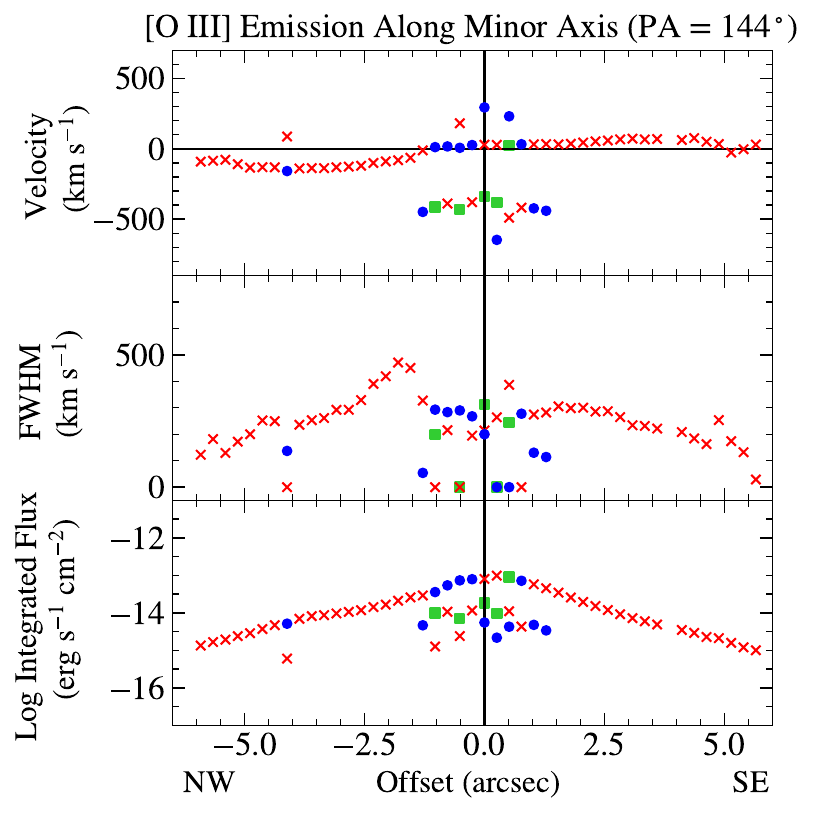}}
{\includegraphics[width=0.425\linewidth]{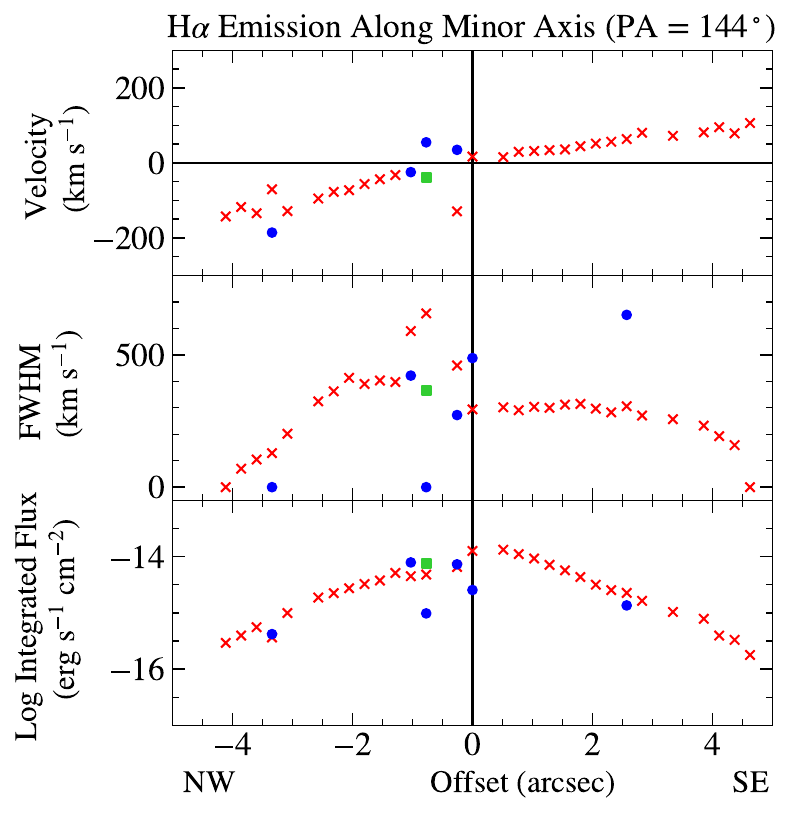}}

 \caption{Continuation of Figure \ref{fig: STIS}, with observed KOSMOS kinematics of [O III] (left) and H$\alpha$ (right) for two PAs.}
\label{fig: KOSMOSobs}
\end{figure*}

\subsection{APO KOSMOS Kinematics}
To get a broader sense of the galactic kinematics, we obtained KOSMOS observations with 2\arcsec\ slit widths. Because KOSMOS has a wide wavelength range, we are able to extract information from both the H$\beta$ $+$ [O~III] and H$\alpha$ $+$ [N~II] emission lines in a single observation. Our kinematic results for the KOSMOS observations are shown in Figure \ref{fig: KOSMOSobs}.

These wider slits exhibit low velocities along the galactic major axis consistent with rotational motion at distances $>$ 1\arcsec -- 2\arcsec, similar to the STIS data, and hints of outflowing motion closer to the AGN. We are unable to detect emission at distances greater than $\sim$7\arcsec~($\sim$1.4 kpc), consistent with the lack of significant star formation, and hence ionized gas, in the greater part of the galaxy \citep{boisson00}.
To compare NGC 3516's rotation and outflow kinematics, we use the rotation curve based on stellar absorption features from \cite{Cherepashchuk2010}.
We measure slightly higher velocities ($\sim$700 km s$^{-1}$) in the ionized gas near the nucleus along the major axis, indicating the effects of outflow even though the components cannot be separated, in contrast to the STIS data.
The outer emission-line velocities follow the projected stellar rotation curve closely, particularly beyond 5\arcsec. Along the host galaxy's minor axis, the stellar velocities should be close to zero, and yet we see emission-line velocities up to 200 km s$^{-1}$ in H$\alpha$ and 500 km s$^{-1}$ in [O~III] near the nucleus, suggesting outflow contributions out to $\sim$5\arcsec. This is consistent with previous kinematics studies of NGC 3516's nucleus, which show somewhat different gas kinematics compared to stellar kinematics at these distances \citep{Cherepashchuk2010, arribas97}.
Overall, we find strong evidence for outflows of ionized gas within $\sim$1\arcsec of the SMBH in NGC~3516, and a mixture of rotation and outflows in the ionized gas up to $\sim$7\arcsec~($\sim$1.4 kpc), with very little ionized gas (by the AGN or stars) at greater distances, as verified in the [O~III] image (See Figure \ref{fig: positionangles}).

 \label{sec:spectra}
\section{Bicone Model of Outflows}
Using these measured kinematics, we constructed a biconical model of outflow for the NLR of NGC 3516. \cite{fischer13} attempted this procedure before, but found ambiguous evidence for outflows in this AGN. This was due to their degenerate model solutions for the inclinations and opening angles of the bicone model, along with insufficient evidence for a symmetrical NLR. By using more observations and various PAs of NGC 3516 to understand the kinematics better, we developed a new model that fits the measured kinematics and explains multiple features seen in the inner parts of NGC 3516.

\subsection{Bicone Model}
\begin{figure*}

\centering  
{\includegraphics[width=0.45\linewidth]{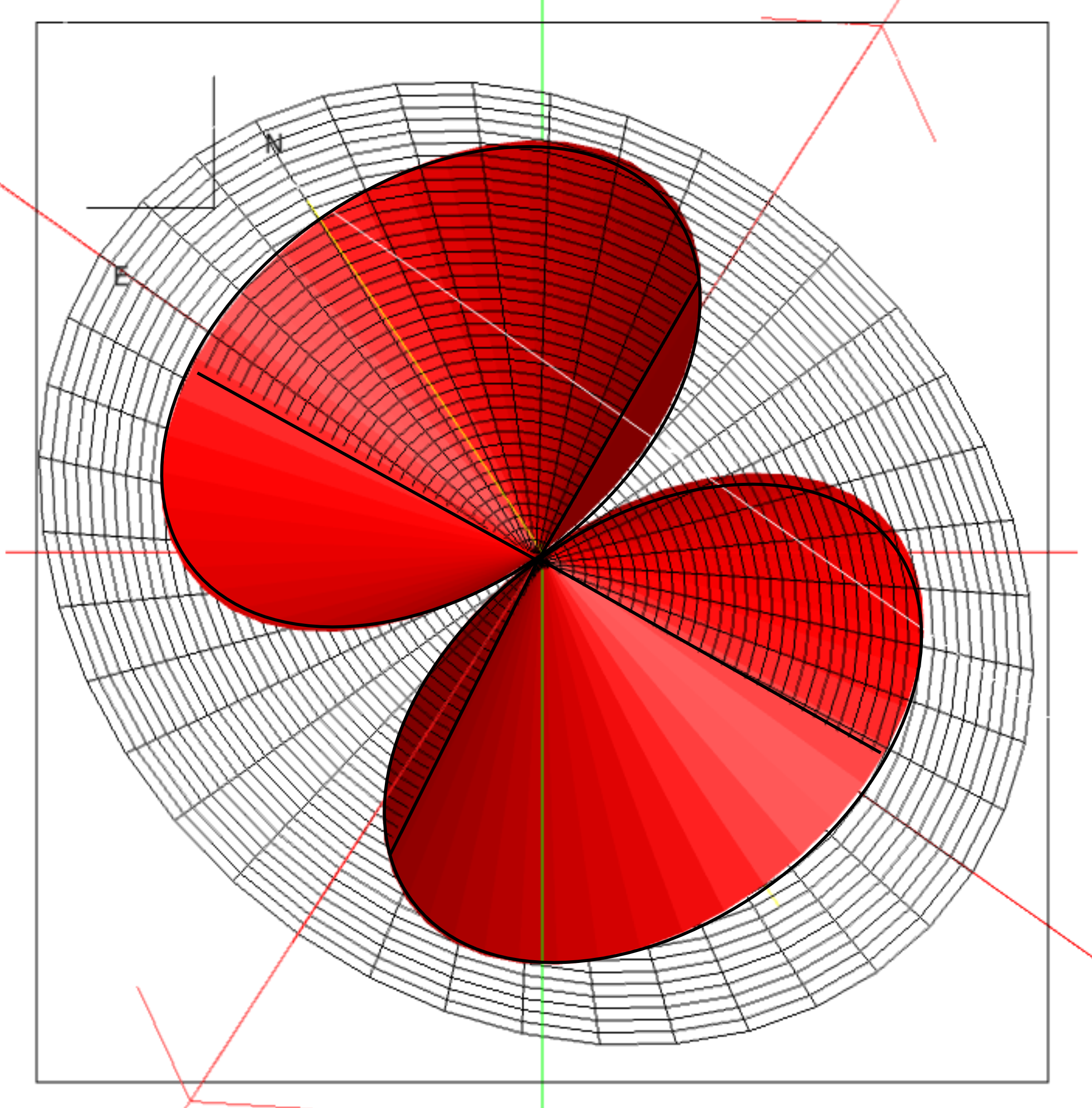}}
{\includegraphics[width=0.45\linewidth]{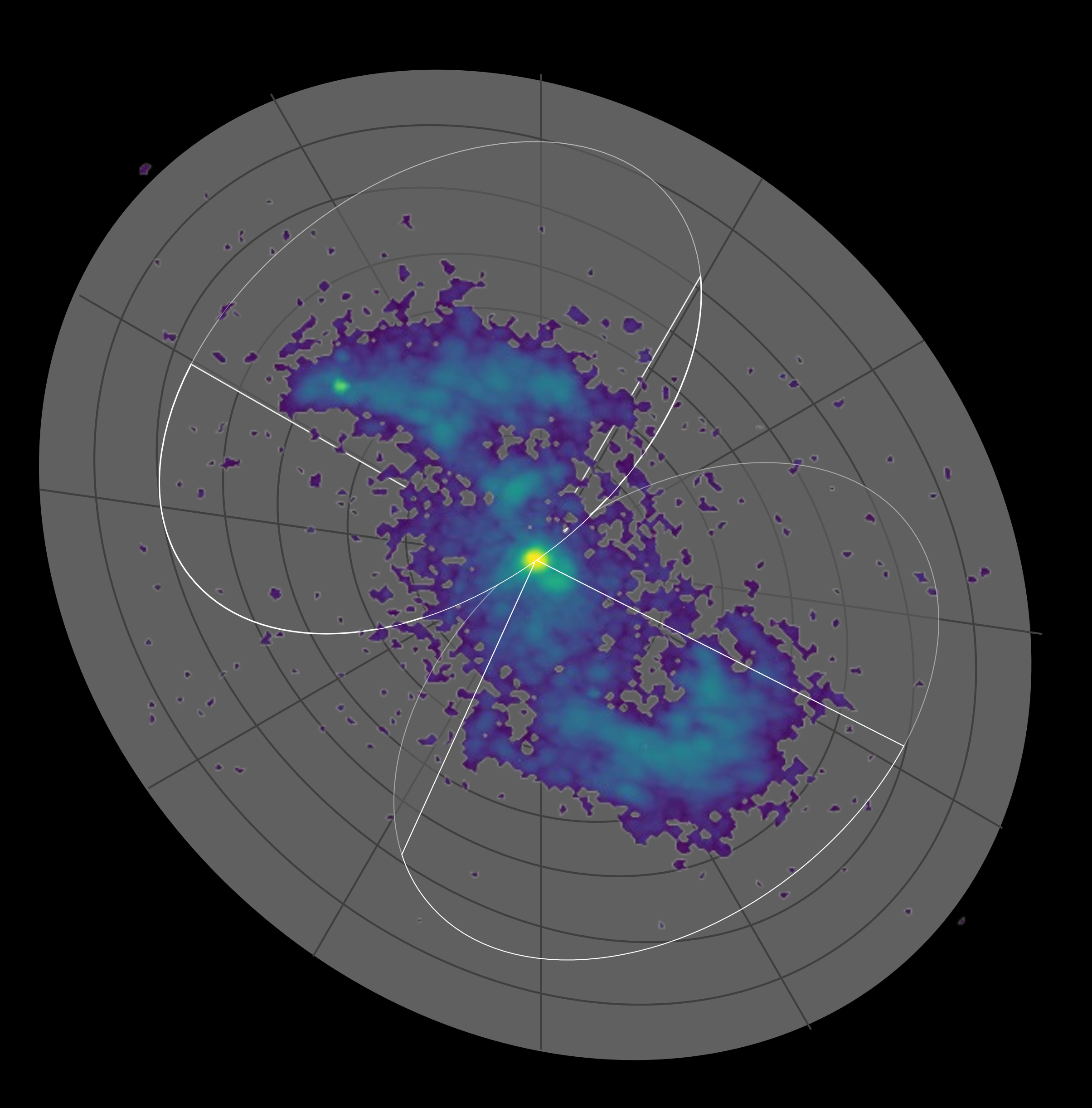}}

 \caption{Visual representation of our bicone model. The length along one cone along its axis is 750 pc. Left: Representation of the bicone at an average half-opening angle of 50\arcdeg\ with the disk of the galaxy through it. Right: Bicone model overlayed on an [O~III] image of NGC 3516 showing the regions of intersection between the ionizing bicone and host galaxy disk. \label{fig:bicone}}

\end{figure*}

The parameters of our bicone model are described in Table~\ref{table:bicone}, and a visual representation is pictured in Figure~\ref{fig:bicone}, for a bicone opening angle of 50\arcdeg. Our bicone model is constrained in a similar way as \cite{fischer13}, based on the kinematic modeling code described in \cite{das05}. To construct a bicone model, we use a symmetric geometry for the NLR, with identical outflowing cones. Several basic parameters are then determined, which constrain both the geometry of the bicone and the velocities along a slit through the galaxy. These parameters include the bicone position angle, which is the angle between north and the axis of the bicone in the plane of the sky; the bicone inclination, which is the angle between the plane of the sky and bicone axis; the inner and outer opening angles, which are measured from the bicone axis to the filled portions of the bicone; the turnover radius, which is the distance from the SMBH at which the NLR clouds stop accelerating and start decelerating; and the maximum height of the bicone, which is the extent of a single cone measured along the bicone axis.

\begin{table}[tt]
\centering
\footnotesize
%\begin{center}
\begin{tabular} 
{|l | c|} 

 \hline

 Parameter & Value  \\
 
 \hline
Galaxy Inclination & 36\degree (NW closer)   \\
Galaxy Position Angle & 56\degree\   \\
\hline
Bicone Position Angle & 35\degree $\pm$ 10\degree\   \\
Bicone Inclination & 40\degree $\pm$ 5\degree(NE closer) \\
Inner Opening Angle & 45\degree $\pm$ 5\degree \\
Outer Opening Angle & 55\degree $\pm$ 5\degree \\
Max. Velocity & 1000 $\pm$ 150 km s$^{-1}$ \\
Turnover Radius & 210 $\pm$ 30 pc  \\
Max. Bicone Height & 750 $\pm$ 100 pc  \\
%STIS & 7403 & O57204010 & 1999 Jan 31 & G750M & 2105 & $6248-6912$ & 0.56 & 0.051 & -137.62 \\

 \hline

\end{tabular}
%\end{center}
\caption{Parameters of the bicone model for NGC 3516. Galaxy inclination and position angle values come from \cite{schmitt00}.}
\label{table:bicone}
\end{table}

In the case of NGC~3516, the morphology of the emission-line gas in the NLR resembles a ``Z" or backwards ``S", similar to that seen in the Seyfert 2 galaxies Mrk~3 \citep{gnilka20} and Mrk~573 \citep{fischer17}, which indicates that we are seeing inner spirals of gas that have been ionized by the AGN. Thus, as noted by the above studies, this gas is located in the intersection between the ionizing bicone and host disk. A further geometric constraint is that our view must be inside or along the edge of the near cone, in order to see the BLR in this Seyfert 1. The kinematic constraints indicate that we are likely looking along the edge, in order to obtain high radial velocities confined to the nucleus at a distance $<$~1\arcsec\ regardless of PA. Thus, to obtain the geometric model in Figure~\ref{fig:bicone}, we varied the PA, inclination, and average opening angle of the bicone until we encompassed the bright NLR emission in the bicone-disk intersection (NW side of the disk is closer, \citealt{ferruit98}). We then varied the opening angles of the bicone, velocity turnover radius, and maximum velocity at turnover, to produce model velocity envelopes that match the majority of the observed radial velocities without encompassing a large amount of empty space in the kinematic velocity plots.

%To generate a model of the outflowing bicone, we vary these parameters and iterate to find a model that both shows the accreting SMBH, as NGC 3516 is a Seyfert 1, and explains the measured velocities with the smallest range of opening angle possible.  

%The north-east cone of the model is tilted towards our LOS, is being viewed along one edge at the nucleus. The disk of the galaxy is tilted towards our LOS in the north-west, and goes through the middle of the NLR bicone. Overlaying the model on top of a picture of the galaxy clearly shows the gas in the S-shaped morphology being ionized by the NLR radiation. 

Figures \ref{fig: STISbiconeshading} and \ref{fig: KOSMOSbiconeshading} show the comparison between our observed radial velocities and model envelopes for each position angle. At STIS PA $=$ 39\arcdeg, near the bicone axis, the overall match is quite good, with the model capturing the high velocity points near the nucleus and most of the observed radial velocities significantly outside of the rotation curve. The amplitudes of some of the redshifted points, particularly at 5\arcsec\ -- 6\arcsec\ SW of the nucleus are not well matched, in both H$\alpha$ and [O~III]. This indicates a possible additional kinematic component not captured by the model. At the other PAs, which are not along the bicone axis, the model slit extractions give large model envelopes, and the observed data points are at least consistent with the envelopes.
%Some model regions have no or few points, likely as a result of extinction by the known dust lanes near the nucleus.

Considering the model versus observed radial velocities in the APO observations, Figure \ref{fig: KOSMOSbiconeshading} shows that the model envelopes are broad, due to the large slit width. The STIS observations in Figure \ref{fig: STISbiconeshading} are also broad at position angles 97$\degree$ and 168$\degree$, though this is due to the projection effects of the velocities from the orientation of the bicone along the slit. The overall trends in the observed radial velocities are well matched by the model, including the capture of a few high-velocity points in [O~III] near the nucleus.

A significant feature of our bicone model is that we are viewing it along one edge, which accounts for the confinement of high radial velocities to the nucleus and previous difficulty in determining a kinematic model of the NLR. 
Interestingly, NGC 3516 has multiple, strong blueshifted absorption features from ionized gas, such as C IV $\lambda\lambda$1548.2, 1550.8, N V $\lambda\lambda$1238.8, 1242.8, and Si IV $\lambda\lambda$1393.8, 1402.8 \citep{Ulrich1983, crenshaw1998, dunn18}, indicating high-column density outflows in our line of sight. This resembles the situation for the well-studied Seyfert 1 galaxy NGC 4151, which also shows very strong outflowing UV and X-ray absorbers \citep{kraemer05, kraemer06, couto16} and a view along the edge of the NLR bicone \citep{das05}.
In both cases, it appears that we are viewing a high-ionization component of the NLR in absorption though the ``filled" portion of the bicone.

One other interesting feature of the model is that the true position and outer opening angles of the bicone as given in Table~\ref{table:bicone} are significantly different from the apparent values, as shown in Figure \ref{fig:bicone}, based on the intersection between the ionizing bicone and the host galaxy disk. The apparent position angle of the [O~III] emission is $\sim20\degree$, while the true position angle is measured as $35\degree \pm 10\degree$. This effect can also be seen in our previous studies of NLR geometries and kinematics, where the ionized gas appears to be offset from the bicone due to how the disk and bicone are oriented and intersect \citep{fischer10, crenshaw10}.

\begin{figure*}
\centering  
{\includegraphics[width=0.45\linewidth]{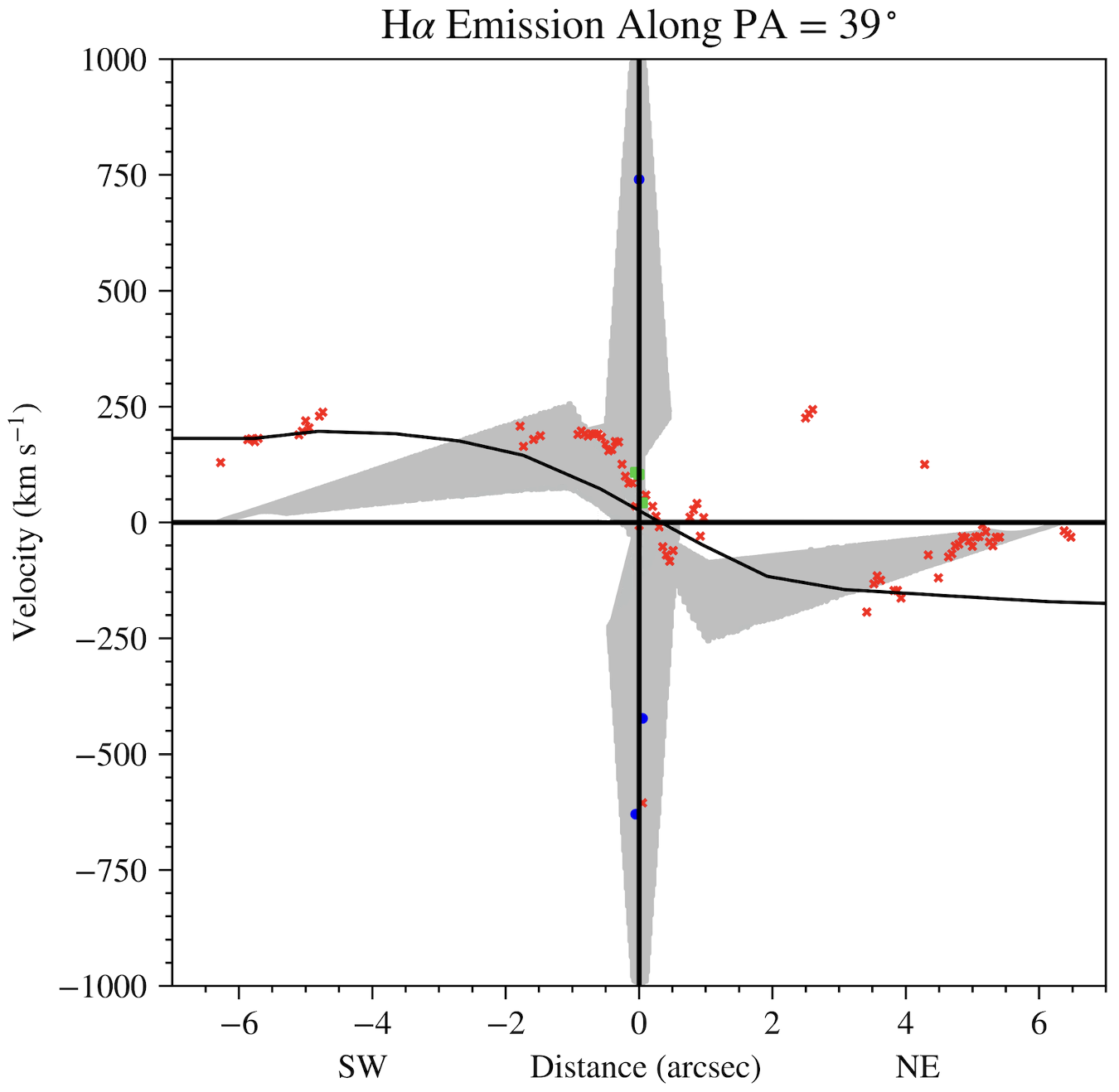}}
{\includegraphics[width=0.45\linewidth]{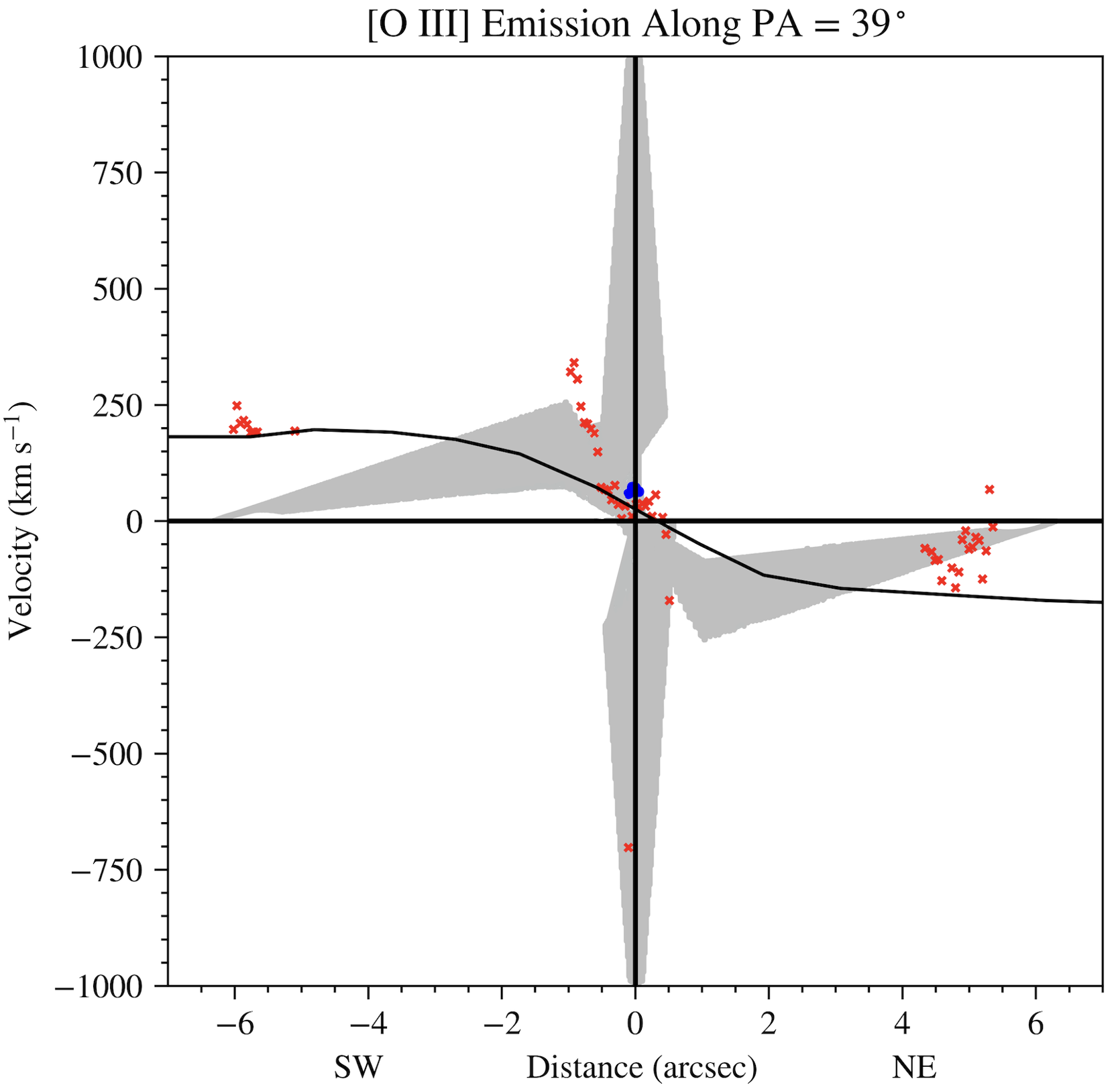}}
{\includegraphics[width=0.45\linewidth]{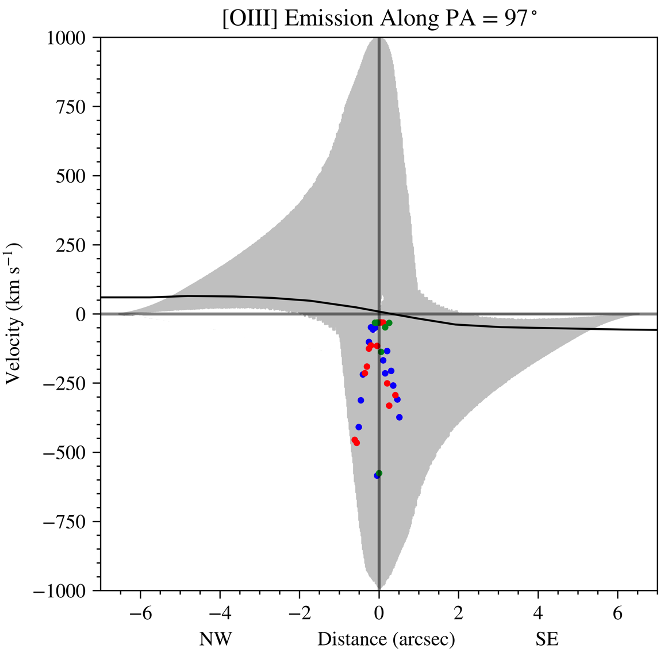}}
{\includegraphics[width=0.45\linewidth]{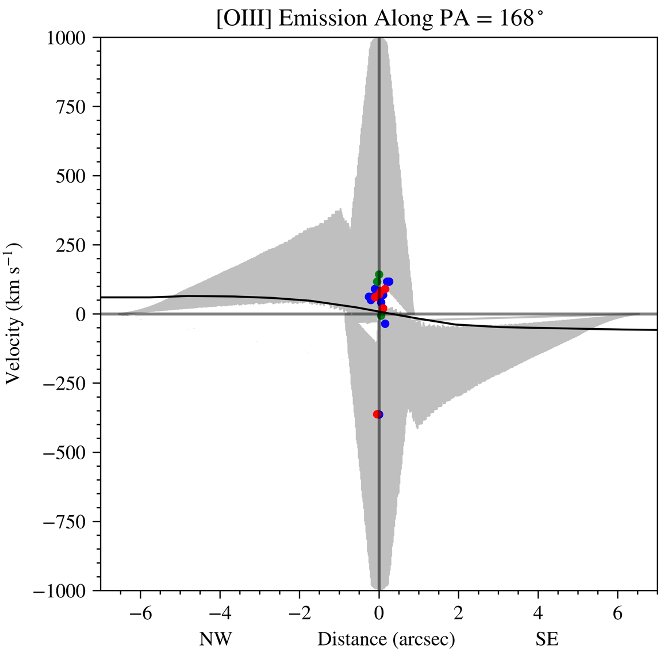}}

 \caption{STIS observed radial velocities (data points) with bicone model envelopes (shading). The continuous black curves are the stellar rotation curves \citep{Cherepashchuk2010} projected along the STIS slit PA. Outflows can be seen as high velocity components seperate from the black rotation curves.}
\label{fig: STISbiconeshading}
\end{figure*}

\begin{figure*}
\centering
{\includegraphics[width=0.45\linewidth]{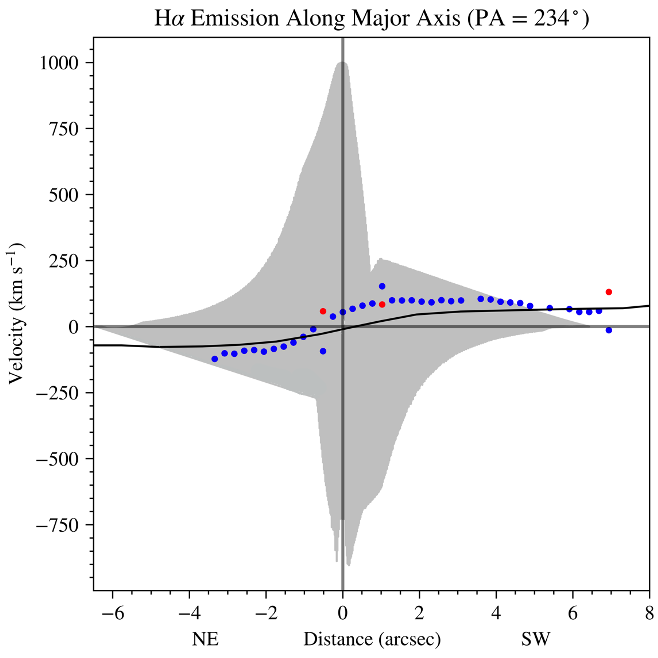}}
{\includegraphics[width=0.45\linewidth]{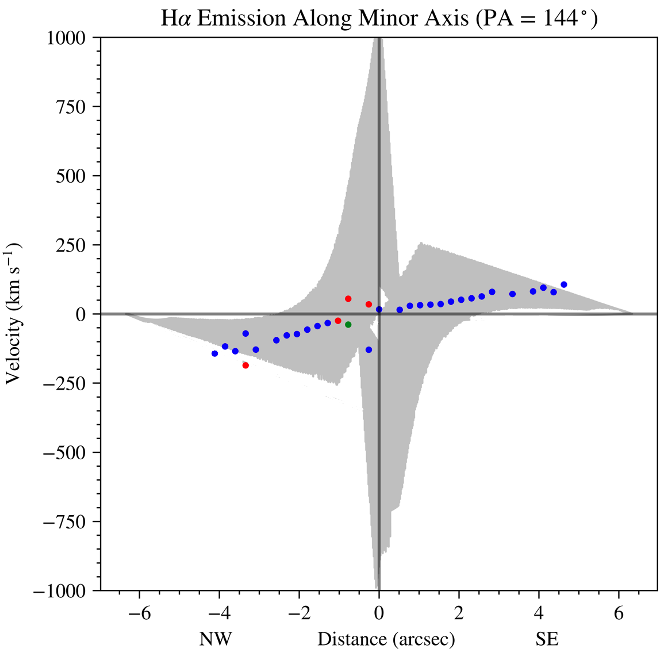}}
{\includegraphics[width=0.45\linewidth]{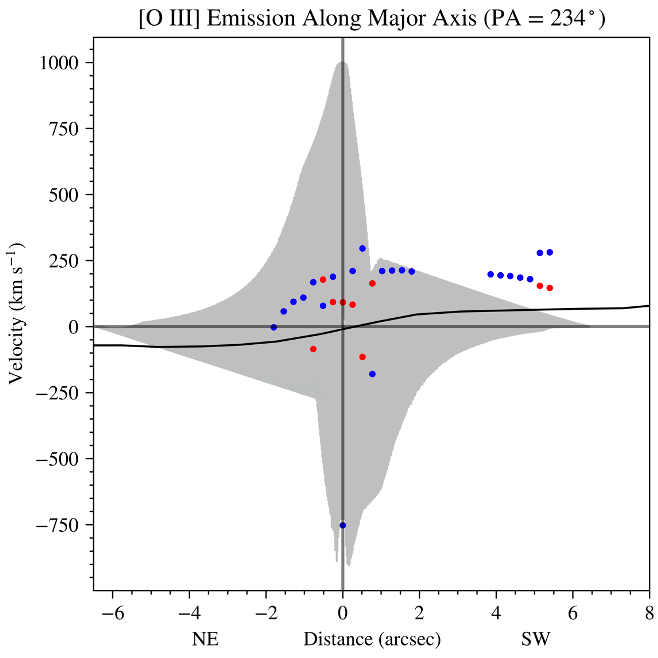}}
{\includegraphics[width=0.45\linewidth]{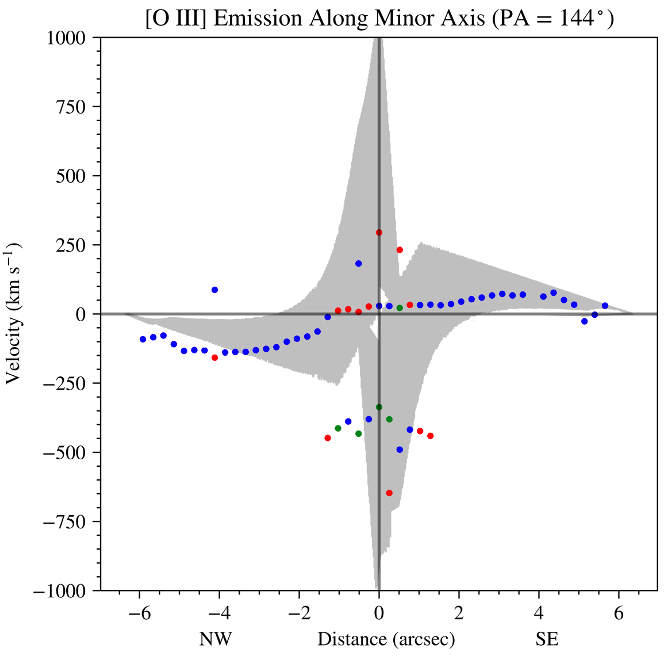}}

 \caption{ Continuation of Figure~\ref{fig: STISbiconeshading}, with observed kinematics from KOSMOS observations.}

\label{fig: KOSMOSbiconeshading}
\end{figure*}
 \label{sec:bicone}
\section{Radiative Driving Analysis}

Recent studies provide strong evidence that NLR outflows in nearby AGN are created in situ from cold gas reservoirs \citep{fischer17, gnilka20} and travel into the galaxy via radiation pressure from the AGN \citep{meena23,falcone24}. The NLR clouds of photoionized gas accelerate from this radiation pressure until the mass of the galaxy and its central SMBH decelerates it \citep{proga00,ramirez12,meena21,falcone24}. By calculating a simple radiative driving model for NGC~3516 from the bolometric luminosity of the AGN and the gravitational potential of the galaxy and SMBH, we can estimate the launch radii of the NLR clouds and determine the turnover distance at which the two forces are equal and the NLR clouds begin to decelerate.

\subsection{Determining Bolometric Luminosity}

Reverberation mapping has revealed a SMBH mass of 2.50 (+0.22 / $-$0.31) $\times$ 10$^{7}$ $M_{\odot}$ in NGC~3516 \citep{derosa18}, corresponding to an Eddington luminosity of $3.15 \times 10^{45}$ ergs s${^{-1}}$. As NGC 3516 hosts a variable AGN, \cite{mehdipour22} reports a bolometric luminosity of 1.2 $\times$ 10$^{44}$ ergs s$^{-1}$ in it's normal high state, and 1.5 $\times$ 10$^{43}$ ergs s$^{-1}$ during changing-look, low phases. This corresponds to an Eddington ratio of 0.038 in the high state to 0.0048 in the low state. The average bolometric luminosity over the NLR can also be calculated from the relation $L_{\mathrm{bol}}=3500 \times L_{5007}$ \citep{heckman04}. \cite{schmitt03} reports an [O~III] luminosity of $10^{41.02}$ erg s${^{-1}}$, corresponding to $L_{\mathrm{bol}}$ $=$ $3.67 \times 10^{44}$ and an Eddington ratio of 0.12. Based on these calculations, it appears that the AGN in NGC~3516 had a significantly higher average luminosity over the past $\sim$5000 years compared to the past few decades. For our analysis, we adopt the bolometric luminosity value of $L_{\mathrm{bol}}$ $=$ $3.67 \times 10^{44}$, which is representative of the long-term radiation pressure on NLR clouds on average.

\subsection{Radiative Driving Model}
To model the radiative driving from the AGN and the gravitational deceleration, we use the equation
\begin{equation}\label{eq: rad driving}
v(r)=\sqrt{\int_{r_1}^{r'}\left[ 4885\frac{L_{44}\mathcal{M}}{r^2}-8.6 \times 10^{-3} \frac{M(r)}{r^2} \right]} dr,
\end{equation}
derived in \cite{das07} and \cite{meena21}. Here, $v(r)$ is the outflow velocity (km s$^{-1}$) as a function of distance~($r$) in parsecs, L$_{44}$ is the bolometric luminosity in units of 10$^{44}$ erg s$^{-1}$, $\mathcal{M}$ is the force multiplier, and $M(r)$ is the mass interior in units of solar masses (M$_{\odot}$). When applied to NLR clouds the only unknown is $r_{1}$, the launch radius, from which the cloud was originally accelerated.

The enclosed mass is calculated from equation A2 in \cite{terzic05},

\begin{equation}
    M(r) = 4\pi\rho_{0}r_{e}^{2}n\kappa^{(3-p)}\Gamma(n(3 - p),Z).\label{enclosedmass}
\end{equation}
where $r_{e}$ is the effective radius, $n$ is the Sérsic index, and $\kappa$ is a constant based on the incomplete gamma function $\Gamma(2n) = 2\gamma(2n, k)$. The values $\rho_0$ and $p$ are given by the equations
\begin{equation}
    \rho_0 = \frac{M}{L}\Sigma_{e}\kappa^{n(1-p)}\frac{\Gamma(2n)}{2r_{e}\Gamma(n(3-p))},
\end{equation}
and
\begin{equation}
    p = 1 - \frac{0.6097}{n} + \frac{0.05563}{n^{2}}.
\end{equation}
For a more complete description and derivation of these quantities, refer to \cite{fischer17,meena21,meena23}.

\begin{figure*}

\centering  
{\includegraphics[width=0.5\linewidth]{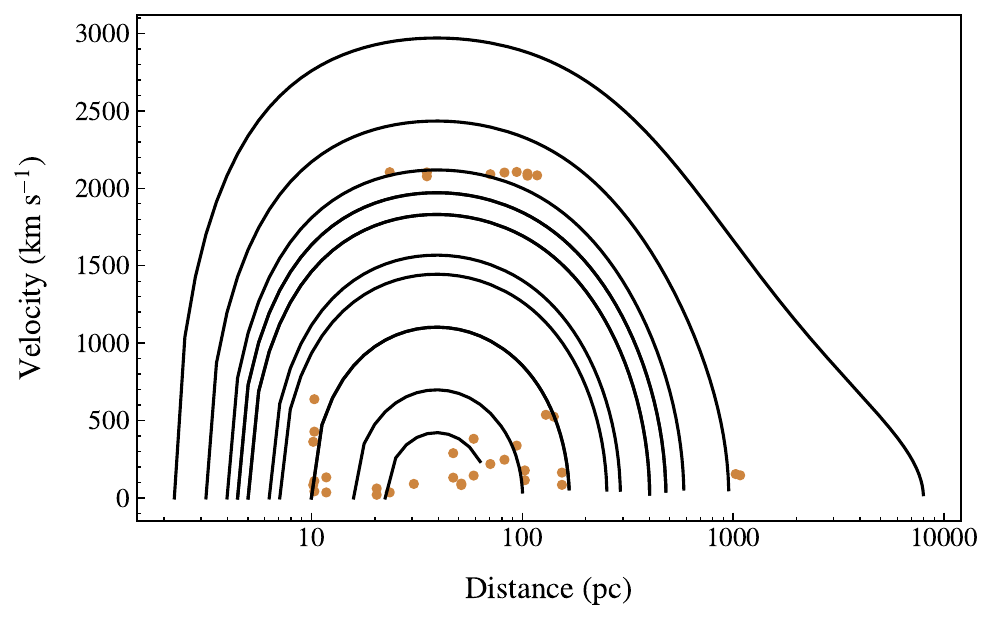}}
{\includegraphics[width=0.45\linewidth]{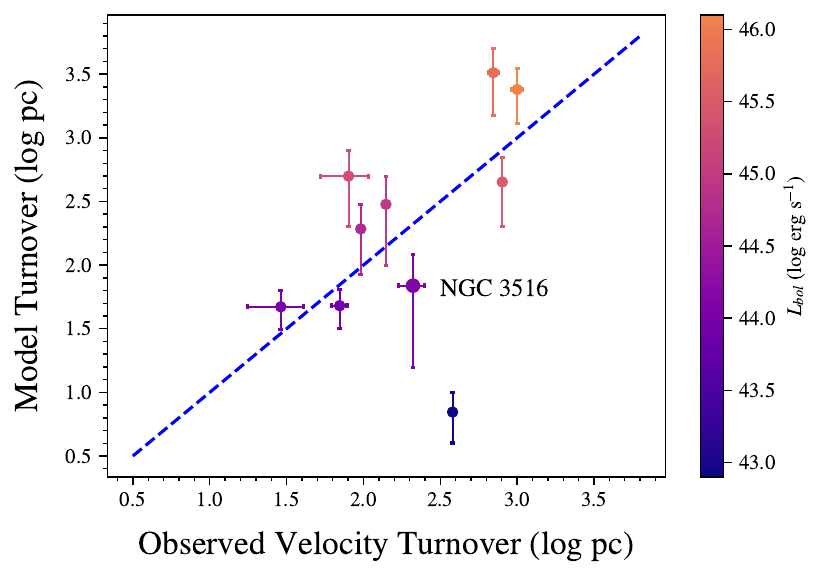}}

 \caption{Left: Trajectories of NLR clouds in NGC~3516 based on our radiative driving and gravitational deceleration models. The model turnover radius in this case is 40 pc, independent of launch radius. The orange points represent deprojected kinematic data from our KOSMOS and STIS observations.
Right: Comparison between observed turnover radii (from fitting kinematics of AGN NLRs with bicone models) and model turnover radii based on the radiative driving plus gravitational deceleration models for 9 AGN spanning a range in luminosity \citep{meena23, falcone24}. The range in the model turnover radius for our added point of NGC~3516 is based on a range in force multiplier $\mathcal{M}$ of 500 - 2000 and adding the uncertainty in converting [O~III] to bolometric luminosity \citep{heckman04}. \label{fig:raddriving}}

\end{figure*}

To determine the enclosed mass with Equation~\ref{enclosedmass}, we use surface brightness fits calculated with GALFIT from \cite{bentz09}, which gives us $r_{e}$, $n$, and $b/a$ for the inner bulge, bulge, and disk components for NGC~3516. Combining this with V, B, and I magnitudes of each component from \cite{bentz18} and the relation between these magnitudes and the M/L ratios \citep{bell01}, we can calculate the trajectories of NLR clouds for a a given initial launch radius. 
These estimated velocities are shown in Figure \ref{fig:raddriving}, with our observed kinematic data overplotted. Our kinematic velocities and radii were de-projected to obtain true values by assuming pure outflow along the disk, where the gas is located, using following equations:

\begin{equation}
        V_{int}(i, \varphi) = \frac{V}{sin(i)sin(\varphi)},
\end{equation}
and
\newline
\begin{equation}
    R_{int}(i, \varphi) = \sqrt{R^2 (cos^{2}(\varphi) + \frac{sin^{2}(\varphi)}{cos^{2}(i)}}),
\end{equation}
where $V_{int}$ is the intrinsic radial velocity, $V$ is the measured velocity, $R_{int}$ is the deprojected spatial scale, $R$ is the measured radius from the nucleus, $i$ is the inclination of the galaxy, and $\varphi$ is the angular offset of the slit from the galaxy major axis \citep{revalski18}.

For this analysis, we adopt benchmark values of $L_{\mathrm{bol}}$ $=$ $3.67 \times 10^{44}$ ergs s$^{-1}$, the average over the NLR, and a force multiplier $\mathcal{M} \approx 1000$, which falls in the middle of the typically used range \citep{meena23, trindade21}. A typical column density of $\log(N_H) = 21.5$ and ionization parameter of $\log(U) = -2.5$ is  assumed. Comparing the turnover radius for NLR clouds from this model and the bicone model in Figure~\ref{fig:bicone}, we find the observed turnover radius (210 $\pm$ 20 pc) to be somewhat higher than our model turnover (40 $\pm$ 15 pc). The right panel of Figure \ref{fig:raddriving} shows this relationship for NGC~3516, alongside other galaxies measured in \cite{meena23, falcone24}. The errors in the model turnover radius come from differing values of the force multiplier $\mathcal{M}$, from 500 to 3000 \citep{trindade22}, along with the intrinsic scatter in the bolometric luminosity correction \citep{heckman04}.

The left panel in Figure \ref{fig:raddriving} shows the NLR cloud trajectories for various launch radii and the deprojected distances and velocities of observed NLR clouds.
For the adopted parameters, the NLR clouds are launched between 4 and 40 pc of the central SMBH, and they can travel up to $\sim$1 kpc. However, we note that NLR clouds at distances $>$ 40 pc and velocities less than $\sim$200 km s$^{-1}$ could be local clouds rotating with the galaxy.
The velocities of clouds at $\sim$2000 km s$^{-1}$ are likely overestimated by a factor of two, due to our assumption of motion along the disk for all clouds, whereas their motions are likely closer to our line of sight near the nucleus as indicated in Figure \ref{fig: STISbiconeshading}. Reducing their velocities to the maximum value of 1000 km s$^{-1}$ alters their launch radii from $\sim$4 pc to $\sim$10 pc., and brings these points much closer to those at lower velocities, indicating clouds from similar launch radii but at different points in their trajectories.

Considering the right panel in Figure~\ref{fig:raddriving}, 
the model and observed turnover values for NGC~3516 agree to within a factor of $\sim$5 as discussed above, similar to many of the other AGN in this plot\footnote{The largest discrepant point is NGC~4051, which likely suffers from resolution effects \citep{meena21, meena23}}.
Thus, NGC 3516 fits within the overall match between model and observed turnover radius, which is driven by increasing luminosity.
In general, this trend indicates that AGN radiative driving and gravitational deceleration by the SMBH and host galaxy are the principal dynamical forces on the NLR clouds.
The effects of luminosity on model turnover radius in AGN outflows are investigated further in M.
K. Shea et al. 2025 (in preparation).
 \label{sec:raddrive}
\section{Conclusions}

We use new KOSMOS and archival STIS observations to derive a biconical model of NLR outflows in NGC 3516, which explains both the kinematics of the ionized emission-line gas and multiple absorption features found in the inner regions of the galaxy. We also construct a model of radiative driving of NLR gas from the central AGN, and compare the observed and turnover radii. Our main conclusions are as follows:
%\newpage
\begin{enumerate}
\item   We detect ionized gas outflows in the nucleus of the galaxy, particularly within 1$\arcsec$ of the central SMBH in NGC 3516. We detect a mixture of outflow and rotation in the galaxy further out, up to a distance of $\sim$7\arcsec~($\sim$1.4) kpc and a maximum measured velocity of $\sim$800 km s$^{-1}$.
\item From our kinematic measurements, we produce a biconical model of the outflows, described in Table \ref{table:bicone}. From this model, we find that one edge of the bicone is directly in our line-of-sight, resulting in the previous difficulty in separating kinematics and likely explaining the presence of multiple components of highly ionized absorption lines in our line of sight.
\item We use these kinematics to derive a model of radiative driving, using the mass of the galaxy and radiation pressure of the AGN. We find a turnover radius that is smaller than the observed turnover radius, although it fits well into the correlation between observed and model turnover radii as a function of luminosity.
\end{enumerate}
 \label{sec:conclusion}

\newpage
\acknowledgments

The authors would like to thank the anonymous referee for the valuable feedback. Many of the data presented in this work are based on observations with the NASA/ESA Hubble Space Telescope and were obtained from the Mikulski Archive for Space Telescopes (MAST), which is operated by the Association of Universities for Research in Astronomy, Incorporated, under NASA contract NAS5-26555. Some of the data used in this article were obtained from MAST, and can be accessed via \dataset[doi:10.17909/5wc0-1w92]{https://doi.org/10.17909/5wc0-1w92}. This research has made use of NASA’s Astrophysics Data System. IRAF is distributed by the National Optical Astronomy Observatories, which are operated by the Association of Universities for Research in Astronomy, Inc., under cooperative agreement with the National Science Foundation. This research has also made use of the NASA/IPAC Extragalactic Database (NED), which is operated by the Jet Propulsion Laboratory, California Institute of Technology, under contract with the National Aeronautics and Space Administration. Some of the observations used in this paper were obtained with the Apache Point Observatory 3.5-meter telescope, which is owned and operated by the Astrophysical Research Consortium.

%\vspace{5mm}
%\facilities{HST(STIS), ...}


%\software{Cloudy \citep{2013RMxAA..49..137F} }

%\appendix

%% the bibliography. The sample63.bib file was populated from ADS.

%\bibliography{bibbo.bib}{}
\bibliographystyle{aasjournal}
\bibliography{bibbo.bib} 
%\nocite{*}

%\appendix
\restartappendixnumbering


\end{document}